# Probing DNA-amyloid interaction and gel formation by active magnetic wire microrheology

Milad Radiom[1,2], Evdokia K. Oikonomou[1], Arnaud Grados[1], Mathieu Receveur[1] and Jean-François Berret[1*]

[1]Université de Paris, CNRS, Matière et systèmes complexes, 75013 Paris, France
[2]Institute for Food, Nutrition and Health, D-HEST, ETH Zürich, Zürich, Switzerland

**Abstract**: Recent studies have shown that bacterial nucleoid-associated proteins (NAPs) can bind to DNA and result in altered structural organization and bridging interactions. Under spontaneous self-assembly, NAPs may also form anisotropic amyloid fibers, whose effects are still more significant on DNA dynamics. To test this hypothesis, microrheology experiments on dispersions of DNA associated with the amyloid terminal domain (CTR) of the bacterial protein Hfq were performed using magnetic rotational spectroscopy (MRS). In this chapter, we survey this microrheology technique based on the remote actuation of magnetic wires embedded in a sample. MRS is interesting as it is easy to implement and does not require complex procedures regarding data treatment. Pertaining to the interaction between DNA and amyloid fibers, it is found that DNA and Hfq-CTR protein dispersion behave like a gel, an outcome that suggests the formation of a network of amyloid fibers cross-linked with the DNA strands. In contrast, the pristine DNA and Hfq-CTR dispersions behave as purely viscous liquids. To broaden the scope of the MRS technique, we include theoretical predictions for the rotation of magnetic wires regarding the generic behaviors of basic rheological models from continuum mechanics, and we list the complex fluids studied by this technique over the past 10 years.

## 1 Introduction

Microrheology aims at studying the mechanical responses of matter to deformations or stresses using a few microliters of sample. The technique is specifically appropriate for biological and costly samples which are available in minute amounts. For liquids, microrheology focuses on determining the static viscosity, whereas for viscoelastic materials it also determines the elastic storage and loss moduli $G'$ and $G''$, respectively. Passive microrheology is based on the thermal diffusion of micron-sized beads, while active microrheology uses externally applied forces to move the beads in the fluid in a controlled fashion. In both passive and active approaches, the monitoring and tracking of the beads are achieved by optical microscopy. From their positions as a function of the time, the displacement and velocity of the beads can be retrieved. The past 20 years have seen increasingly rapid advances in the field of microrheology. In general, bead microrheology relies on complex procedures regarding statistical data treatments, such as the transformation of the bead position into the mean-squared displacement (MSD), or using models to convert the MSD into $G'(\omega)$ and $G''(\omega)$ as a function of the frequency $\omega$ [1,2]. In many practical cases, it may be useful to develop alternative microrheology methods, especially methods that do not require lengthy data processing.

Recently, we have combined a novel type of microrheology probes with the technique of rotational magnetic spectroscopy (MRS). The combination meets the requirements for a method that is easy to implement and accessible to a wide range of laboratories. The probes that were developed in our group are anisotropic magnetic wires of length between 1 and 100 µm, and diameters between 200 nm and 2 µm. Under the application of a rotating magnetic field, the wires behave like micro-rheometers embedded in the fluid. Direct analogies with standard rotational rheometry experiments such as creep-recovery and steady shear experiments have been put forward, allowing comparison between the two techniques [3]. In MRS, it is found that with increasing frequency, a hydrodynamic transition between steady and hindered motion of the wire occurs and that the transition frequency scales with the inverse of the fluid static viscosity. For viscoelastic samples, the high-frequency oscillation amplitude is found to scale with the inverse elastic modulus $G'$. Therefore, the determination of viscosity and elasticity is achieved by simple measurements of the wire orientation. This type of approach offers multiple possibilities to determine the rheological properties of fluids and of soft solids, in static as well as in dynamic regimes. In what follows, soft solids will refer to elastic materials characterized by a storage modulus greater than the loss modulus and by a yield stress.

Recently, this approach has shown interesting results pertaining to the coupling of bacterial proteins and DNA molecules [4]. In living organisms, it is well-known that genetic material is in a crowded state. In eukaryotic cells, the volume fraction *i.e.* the volume occupied by macromolecules relative to the volume of the compartment is estimated at 30%, whereas it can reach ~ 50 % in prokaryotes [5]. The bacterial nucleoid, *i.e.* the region containing the genetic material, is usually submicron, while the size of the DNA outline is around 1 mm [6]. DNA folding is facilitated by a multitude of biophysical interactions, which are only partially understood. For the last decade, a wide range of techniques have been used to study the binding of nucleoid associated proteins (NAPs) with DNA strands, including bacterial amyloids such as Hfq. These technologies include fluorescence microscopy imaging of single DNA molecules confined inside nanofluidic channels [7-10], atomic force microscopy [8,11,12], small angle neutron scattering [9], synchrotron radiation circular dichroism [12-14] and isothermal titration [11]. These studies have suggested that NAPs bind to DNA through electrostatic interaction, leading to modification of mechanical properties, structural organization, and/or bridging interactions between different segments of the same or different DNA molecules [15]. These protein complexes may in turn induce DNA cross-linking and entanglements, together with a slowing down of molecular machinery and eventually an increase of viscoelasticity within the nucleoid. Larger Hfq-mers may also form anisotropic amyloid fibers, which have even more significant effects on DNA dynamics. To test this hypothesis, microrheology experiments on DNA dispersions with the amyloid terminal domain of Hfq were performed with the MRS technique [4]. In this chapter, we survey the microrheology technique discussed previously. We also show how it allowed us to demonstrate the existence of gel between protein aggregates of amyloid type and DNA strands.

## 2 Materials

Prepare all the colloidal dispersions using ultrapure MilliQ water with a resistivity of 18.2 MΩ cm and (unless otherwise specified) a pH of 5.5 at T = 25 °C.

### 2.1 DNA and proteins

1. (dAdT)59 oligonucleotide bearing sixty-nine Adenine-Thymine (AT) base pairs. Prepare stock solutions at the base pair molar concentrations of 1.8 and 7.8 mM. The (dAdT)59 dispersions are transparent and exhibit liquid-like behavior.
2. Hfq-CTR peptide corresponding to the amyloid CTR terminal domain of the bacterial nucleoid-associated protein Hfq: prepare a 20 g L$^{-1}$ stock solution. The Hfq-CTR dispersion is transparent and exhibits liquid-like behavior (*see* **Note 1**).
3. Hfq-CTR mutant peptide does not form amyloid fibers and does not bind to DNA [4].
4. Phosphate Buffer Saline (PBS) at pH 7.4

### 2.2 Chemicals for the MRS experiment

1. Poly(diallyldimethylammonium chloride) (PDADMAC) of molecular weight $M_w$ < 100000 kDa. The PADAMAC stock solution is slightly yellow and viscous and has a concentration of 35% by weight.
2. Maghemite (γ-Fe$_2$O$_3$) nanoparticles coated with poly(acrylic acid) of molecular weight 2 kDa: prepare a 1 g L$^{-1}$ aqueous dispersion. The particles must have a diameter around 10 nm and a saturation magnetization $m_S$ of 3.5×10$^5$ A m$^{-1}$ (or 72 emu g$^{-1}$ in cgs units). The coated particles, γ-Fe$_2$O$_3$@PAA$_{2K}$ are obtained by precipitation-redispersion [16,17].
3. Ammonium chloride salt NH$_4$Cl white powder (*see* **Note 2**).
4. Dialysis cassette with membranes of molecular weight cutoff of 10000 Da.
5. Glycerol. Water-glycerol mixtures must be used for calibrating the wire magnetic properties. Prepare three water-glycerol samples at weight fractions 0.498, 0.776 and 0.936, corresponding to static viscosities of 4.97, 35.5 and 270.9 mPa s at room temperature ($T$ = 25 °C) [18].

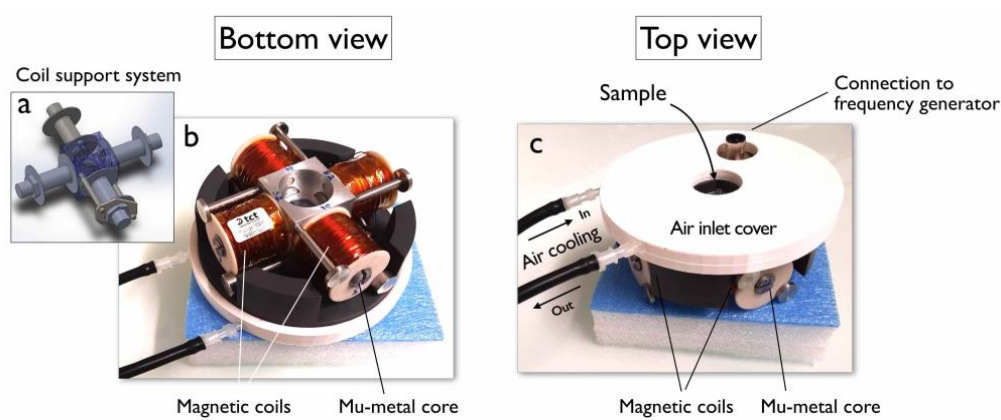

**Figure 1**: Homemade magnetic rotating field device. a) Aluminum and poly(vinyl chloride) support structure on which the 4 coils of resistance 4.68 Ω are mounted. b) Bottom view of the magnetic rotating field device, showing the coils and the mu-metal cores. Each pair of coils produced a magnetic field from 0 to 30 mT at the sample location. c) Top view of the magnetic rotating field device, showing the sample location, the air-cooling inlet and outlet and the air inlet cover.

**2.3 Magnetic rotating field device and equipment**

The rotating magnetic field is produced by a homemade device composed of four coils consisting of 745 turns (wire diameter 0.5 mm) on 17 layers and oriented at 90° to each other, as shown in Fig. 1 (*see* **Note 3**). The core of the coils is made of 80:20 nickel-iron alloy (mu-metal), which is a soft ferromagnetic material with high permeability. The mu-metal cores are characterized by a saturation magnetization of 0.7 T and a magnetic coercivity inferior to 2 A m$^{-1}$. Each pair of aligned coils can produce a magnetic field between 0 and 30 mT at the sample location. The electronic signal in the coils (resistance of 4.68 Ω) is produced by a low frequency generator coupled to a homemade current amplifier. To produce a rotating magnetic field, the electrical signal entering each pair of coils is phase-shifted by 90°. A stream of air directed toward the measuring cell through an air inlet cover is used to thermalize the sample between room temperature and 50 °C (*see* **Note 4**).

**2.4 Optical microscope and environment**

1. Inverted microscop equipped with ×20, ×40, ×60 and ×100 objectives allowing bright field and phase contrast measurements. The numerical apertures of the objectives are 0.4, 0.55, 0.7 and 1.3 respectively (Fig. 2).
2. QImaging EXi Blue camera and Metamorph software are used as acquisition system.
3. Homemade magnetic rotating field device (Fig. 1) and equipment comprising an aluminum and poly(vinyl chloride) support structure on which the coils are mounted and an air inlet cover for the sample thermalization.
4. Function generator with arbitrary signals, 50 MHz pulses and 2 channels (*see* **Note 5**).
5. Homemade current amplifier.
6. 100 MHz 2-channel oscilloscope.

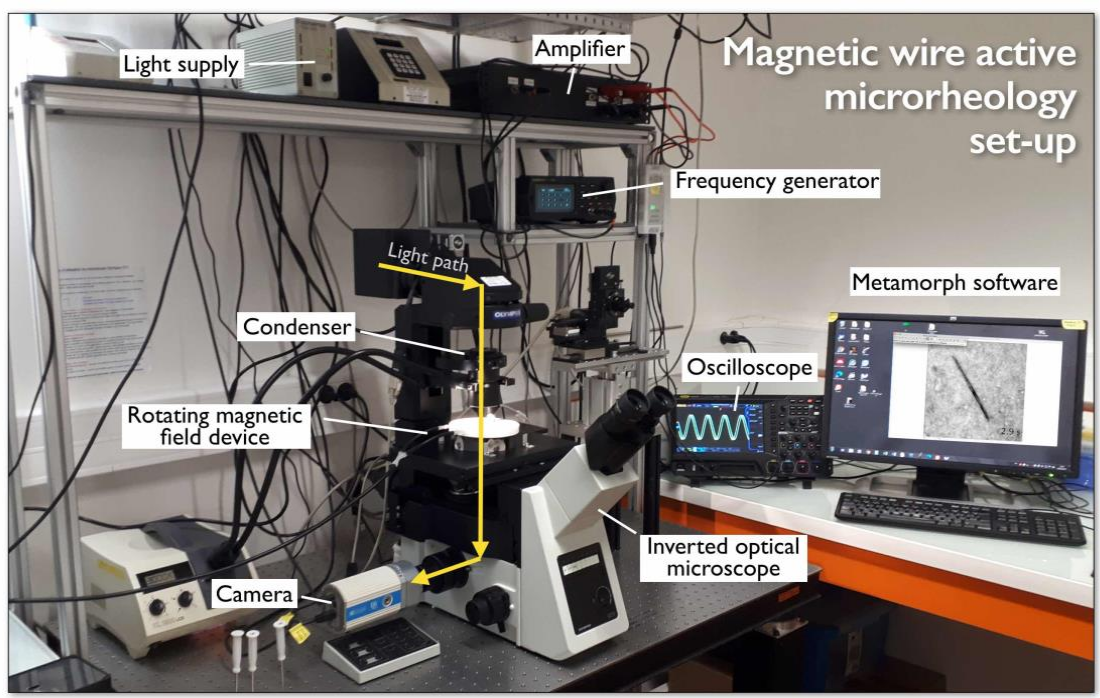

*Figure 2*: *Optical microscopy set-up used to perform magnetic rotational spectroscopy (MRS) experiments.*

### 2.5 Instrument and Accessories
1 - Sonicator bath working at the frequency of 40 kHz and an applied power of 110 W. The sonicator is used to shorten the length of the wires from tens of micrometers to a few micrometers, which is required for some specific environments [19].
2 – Gene Frame dual adhesive system (*see* **Note 6**).

### 2.6 Analysis tracking softwares
Images of wires are digitized and treated by the ImageJ software and related plugins (http://rsbweb.nih.gov/ij/, *see* **Note 7**).

## 3 Methods
### 3.1 Magnetic wires as viscosity sensors
The wires are fabricated using an electrostatic co-assembly process between γ-$Fe_2O_3$@$PAA_{2K}$ and PDADMAC, the cationic polymers acting as the "glue" for the nanoparticles [20]. The bottom-up approach is based on the desalting transition of mixed solutions using a dialysis cassette of cutoff 10000 Da at pH8. Unidirectional growth of the assembly is induced by a 0.3 Tesla magnetic field applied during dialysis. Fig. 3 displays optical (a) and scanning electron microscopy (b-d) images of magnetic wires deposited on a glass substrate, from which the length and diameter distributions are derived, respectively. For the wire sample shown, the median length and diameter are $L$ = 26.9 ± 13.9 (SD) μm and $D$ = 1.21 ± 0.32 (SD) μm. The wire length distribution will be exploited in the data analysis section (Section 3.1). Of note, the wires have inherited the superparamagnetic characteristics of the γ-$Fe_2O_3$@$PAA_{2K}$ particles, which means that they acquire a macroscopic magnetic moment only when placed in a constant magnetic field, a feature that is decisive in modeling their rotation [21].

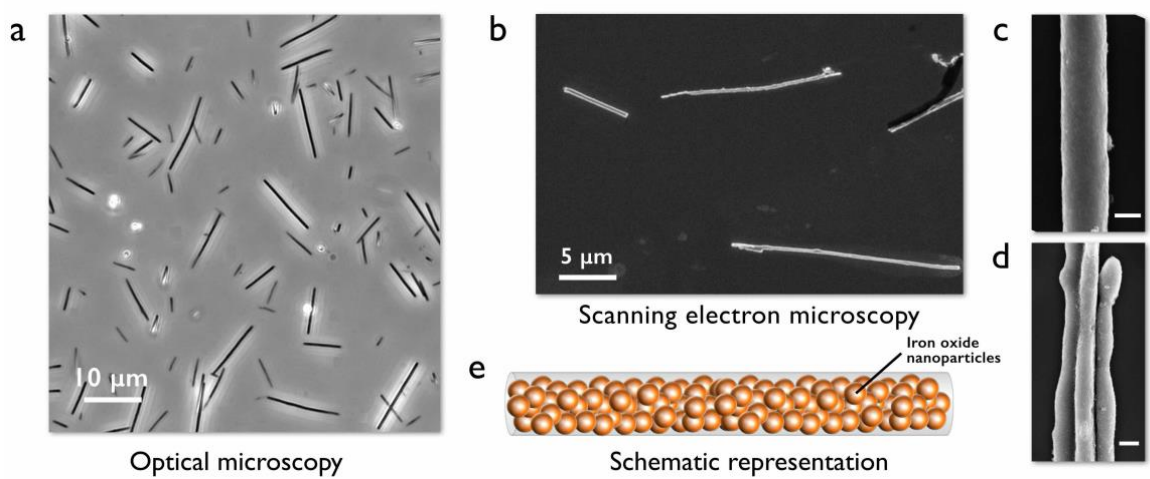

**Figure 3**: a) Phase-contrast image of magnetic wires deposited on a glass substrate and observed by optical microscopy (60×). For these wires, 6.8 nm γ-*Fe₂O₃* particles were used [43]. b-d) Scanning electron microscopy (SEM) images of magnetic wires at various magnifications. The bars in c) and d) are 200 and 300 nm respectively. e) Schematic representation of a magnetic wire made of iron oxide nanoparticles.

### 3.2 Theory of Magnetic Rotational Spectroscopy (MRS)

With the MRS technique, a rotating magnetic field $B$ is applied at the angular frequency $\omega$ to the sample containing micron-sized wires, and the movement of the wires is monitored by time-lapse optical microscopy. The wire is submitted to a magnetic torque $\Gamma_{Mag}(B)$ of the form [21,22]:

$$\Gamma_{Mag}(B) = \frac{\chi^2}{2\mu_0(2+\chi)} V B^2 \sin 2(\omega t - \theta(t)) \quad (1)$$

where $\chi$ is the wire magnetic susceptibility, $\mu_0$ the vacuum permeability and $V = \pi D^2 L/4$ the volume of the wire. In the sinus argument, $\theta$ describes the wire orientation and $\omega t - \theta(t)$ the angle of the wire with respect to the field (Fig. 4a). Immersed in a medium with viscoelastic properties, the wire experiences a viscous and an elastic torque that both impede its rotation. As shown below, the viscosity and elasticity have different effects on the wire rotation.

### 3.1.1 Effect of viscosity on wire rotation

Fig. 4b-f displays the generic behavior derived from the mechanical constitutive equation of a viscous Newton liquid of static shear viscosity $\eta$ [21,23]. At low frequency (Fig. 4b), the wire rotates in phase with the field; the motion is synchronous and $\theta(t) = \omega t$. With increasing $\omega$, the friction torque increases and above a critical value noted $\omega_C$ the wire can no longer follow the field. The wire then undergoes a transition between a synchronous and an asynchronous regime at [21,22]:

$$\omega_C = \frac{3}{8\mu_0} \frac{\Delta\chi}{\eta} g\left(\frac{L}{D}\right) \frac{D^2}{L^2} B^2 \quad (2)$$

In the previous equation, $\Delta\chi = \chi^2/(2+\chi)$ denotes the anisotropy of susceptibility between parallel (*i.e.* along the main axis of the wire) and perpendicular directions and $g(x) = ln(x) - 0.662 + 0.917x - 0.050x^2$ [24]. For data processing, it is convenient to combine the geometric characteristics of the wire in Eq. 1 into the dimensionless parameter $L^* = L/D\sqrt{g(L/D)}$, leading to:

$$\omega_C = \frac{3}{8\mu_0} \frac{\Delta\chi}{\eta} \frac{B^2}{L^{*2}} \quad (3)$$

As shown in Fig. 4c and 4d, above $\omega_C$, the wire is animated by a back-and-forth motion and the angle $\theta(t)$ displays oscillations. The transient behaviors $\theta(t)$ presented intend to highlight two important quantities here: the average angular velocity $\Omega = d\theta/dt$ (straight line in Fig. 4e) and the angle $\theta_B$ by which the wire returns after a period of increase (Fig. 4f).

With MRS, there are several methods to measure the viscosity of a fluid. The simplest one is to experimentally determine the critical frequency $\omega_C$ and to retrieve $\eta$ from Eq. 1. By repeating the measurement of $\omega_C$ on wires of different lengths, it is then possible to verify the $L^{*-2}$-dependence in Eq. 2, which provides strong support for the validity of the model. This later protocol significantly increases the accuracy of the viscosity measurement [21,25]. Another method is to calculate the average angular velocity $\Omega(\omega)$ or the angle $\theta_B(\omega)$ from the $\theta(t)$- traces and to compare them with constitutive model predictions. For Newton and Maxwell fluids [22,25], it is found that:

$$\begin{aligned} \omega \leq \omega_C \quad & \Omega(\omega) = \omega \\ \omega \geq \omega_C \quad & \Omega(\omega) = \omega - \sqrt{\omega^2 - \omega_C^2} \end{aligned} \quad (4)$$

In Eq. 3, $\Omega(\omega)$ increases linearly with the frequency and passes through a maximum at $\omega_C$ before decreasing (Fig. 4e). The probe response over a broad spectral range appears as a resonance peak similar to that found in mechanical systems [26,27]. Pertaining to the return angle $\theta_B(\omega)$, the angle is found to decrease with increasing frequency in accordance with the Newton constitutive equation prediction [21,22]. Away from the transition ($\omega \gg \omega_C$), the decrease goes approximately as $\theta_B(\omega) \sim \omega^{-1}$ (Fig. 4f). Again, methods based on frequency analysis further increase the accuracy of the viscosity measurement.

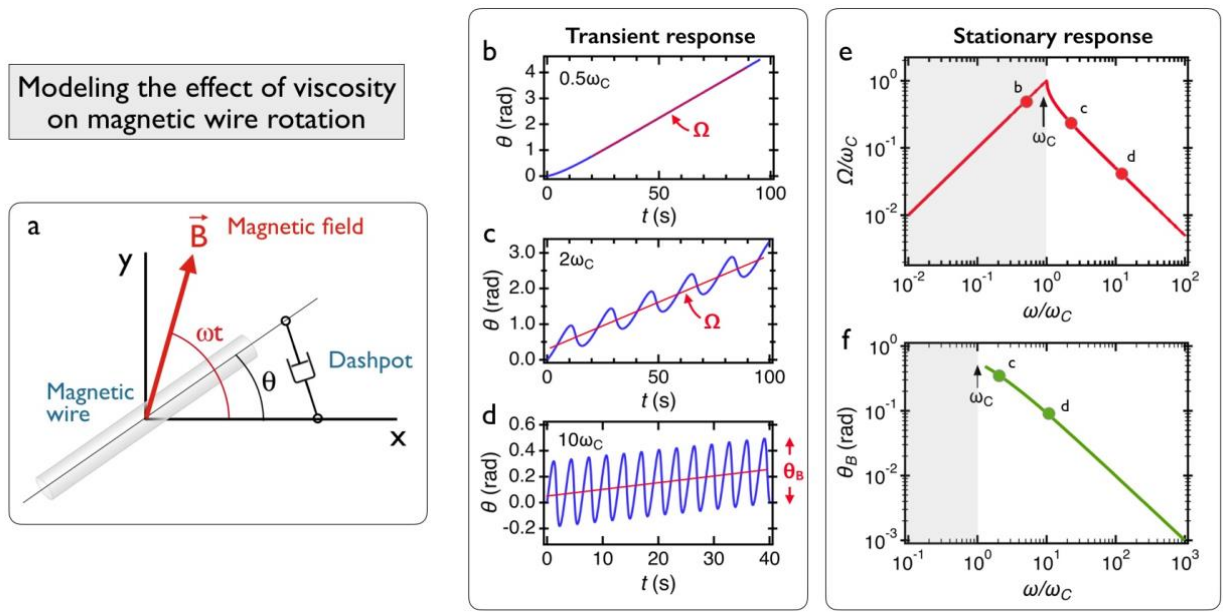

*Figure 4*: a) Schematic representation of a wire actuated in a Newton viscous liquid. The model is represented by a dashpot. b-d) Time dependences of the wire orientation $\theta(t)$ calculated for the Newton model outside the critical frequency $\omega_C$. The straight lines in red denote the average rotational velocity $\Omega$. e,f) Newton predictions for the average angular frequency $\Omega(\omega)$ and for the return angle $\theta_B(\omega)$. The grey area indicates the synchronous regime.

### 3.1.2 Effect of elasticity on wire rotation

Soft solids are materials characterized by a yield stress $\sigma_Y$, i.e. these materials flow only if a stress greater than $\sigma_Y$ is applied. For applied stresses less than $\sigma_Y$, the material is submitted to a deformation only. For such materials, the elastic modulus $G'(\omega)$ is found to be larger than the loss modulus $G''(\omega)$ on a broad frequency range [28]. The rheological models generally used to describe soft solids are the Kelvin-Voigt and the Standard Linear Solid (SLS) models. The Kelvin-Voigt model is accurate for modeling creep experiments, but is not appropriate to describe transient responses, e.g. in controlled deformation experiments [3]. The SLS model combines a Maxwell element of viscosity $\eta$ and elasticity $G$ and a Hookean spring of elasticity $G_{eq}$ in parallel (Fig. 5). The solution of the corresponding constitutive equation [23] shows that regardless of the applied angular frequency $\omega$, the average rotational velocity $\Omega(\omega)$ is identically zero. This behavior is illustrated in Figs. 5b and 5c that display the rotation angle $\theta(t)$ at two frequencies. The equality $\Omega(\omega) = 0$ also means that for such materials, there is no synchronous rotation regime for the wire rotation, and that the critical frequency $\omega_C = 0$, in agreement with the property that soft solids are characterized by infinite static shear viscosity [28].

Fig. 5d displays prediction for the oscillation amplitudes $\theta_B(\omega)$ versus $\omega$. The asymptotic values of the back-and-forth angle at low and high frequency are given by [23]:

$$\lim_{\omega \to 0} \theta_B(\omega) = \theta_{eq} \quad and \quad \lim_{\omega \to \infty} \theta_B(\omega) = \frac{\theta_0 \theta_{eq}}{\theta_0 + \theta_{eq}} \quad (5a)$$

$$\text{where } \theta_{eq} = 3\Delta\chi B^2/4\mu_0 G_{eq} L^{*2} \text{ and } \theta_0 = 3\Delta\chi B^2/4\mu_0 G L^{*2} \quad (5b)$$

In Eqs. 5, $G_{eq}$ and $G$ denote the equilibrium and static storage elastic moduli, respectively. The curves shown in Fig. 5d are for $\theta_0$ = 0.01, 0.1 and 1, and $\theta_{eq}$ = 1. A major difference between Newton viscous liquids and soft solids regarding the actuated wire motion can be summarized as follows: the high frequency limit of the oscillation angle $\theta_B$ is null in the former and finite in the later, its value scaling with the inverse of the elastic modulus $G$. These results show that in addition to viscosity measurements, MRS can also determine the elasticity properties of soft materials.

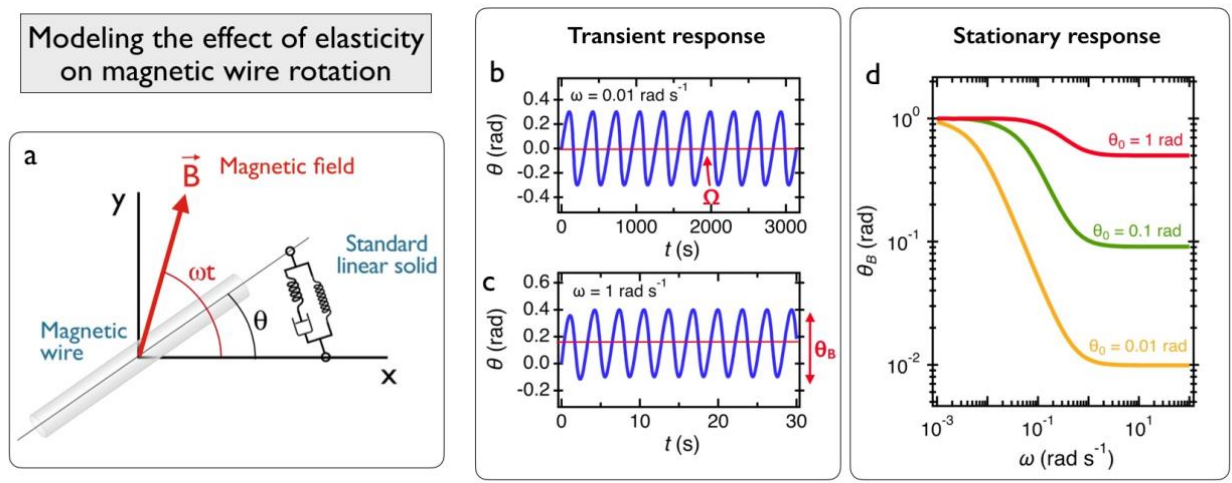

***Figure 5***: *a) Schematic representation of a wire actuated in a Standard Linear Solid model [3,23]. The model is represented by a spring in parallel with a Maxwell element (a dashpot and a spring in series). b-c) Time dependences of the wire orientation $\theta(t)$ calculated for the Standard Linear Solid model at two different frequencies. The straight lines in red denote the average rotational velocity $\Omega$, which is null. d) Standard Linear Solid predictions for the oscillation angle $\theta_B(\omega)$ as a function of the frequency for three values of $\theta_0$ (using $\theta_{eq}$ = 1 rad). The characteristic of viscoelastic materials is that $\theta_B(\omega)$ exhibits a non-zero limit at high angular frequency.*

### 3.1.3 Generic behaviors of basic rheological models

There are four basic models to describe the linear rheological properties of complex fluids and materials. These models are those of Newton, Maxwell, Kevin-Voigt (or its more accurate version, the Standard Linear Solid) and Hooke. These fluids and materials are described as a dashpot, a dashpot and a spring in series, a dashpot and a spring in parallel and a spring, respectively. Table 1 summarizes the generic behaviors of a magnetic wire subjected to a rotating field as a function of angular frequency $\omega$ when placed in one of the four basic rheological models. The predictions are obtained by solving the constitutive equations taking into account that the torques applied to the wire can be of viscous and elastic origin [23]. As can be seen from Table 1, the signatures of each model with respect to the motion of a rotating wire are straightforward, especially with respect to the critical frequency or the oscillation angles at high frequencies. This allows us to conclude that the MRS method performed using superparamagnetic wires at magnetic fields in the milli-Tesla range can unambiguously differentiate between a viscous fluid and a soft solid with a yield stress.

| Rheological model | Static shear viscosity | Yield stress | Sync./Async. transition in MRS | $\Omega(\omega)$ | $\lim_{\omega \to \infty} \theta_B(\omega)$ |
|---|---|---|---|---|---|
| Newton | $\eta$ | No | Yes | Eq. 3 | 0 |
| Maxwell | $\eta = G\tau$ | No | Yes | Eq. 3 | $\theta_0$ |
| Standard Linear Solid | infinite | Yes | No | 0 | $\theta_0 \theta_{eq}/(\theta_0 + \theta_{eq})$ |
| Hooke | infinite | Yes | No | 0 | $\theta_{eq}$ |

**Table 1**: List of models used in rheology to describe the linear mechanical response of fluids and soft materials, as well as the predictions for the wire rotation in the different regimes.

### 3.1.4 List of samples already studied with Magnetic Rotational Spectroscopy

During the past years, our group has studied a number of complex fluids with the MRS method. These fluids are to date 7 in number (Table 2), and they comprise water-glycerol mixtures [18,21,29], surfactant wormlike micelles [30], polysaccharide gels [23], the cytoplasm of living mammalian cells [19,25], a pulmonary surfactant mimetic having been exposed to nanoparticles [31], human pulmonary mucus [32] and the system discussed in this chapter, DNA and Hfq-CTR proteins [4]. In all these cases, a complete description of the wire rotation has been provided, and was found to agree with the continuum mechanics predictions of Table 1. In addition, in four of these seven cases, standard rheometry measurements (*i.e.* using a rheometer and a cone-and-plate tool for shearing) have allowed verifying that the viscosity or elastic modulus values retrieved from MRS were accurate (*see* **Note 8**). Results obtained by other research groups using the MRS technique have been reported in the literature on a wide variety of samples including hydrogels [33], butterfly saliva [34], ceramic precursors [35], nanodroplets and thin films of various liquids [36,37], water-glycerol [38] and blood clot [39].

### 3.2 Wire calibration using fluids of known viscosity

The wire magnetic calibration consists in determining the anisotropy of the magnetic susceptibility $\Delta\chi$ between parallel and perpendicular directions (Eq. 3). To this aim, MRS experiments are performed on a series of water-glycerol mixtures of viscosities 5 to 270 mPa s (*see* **Note 9**) [18]. Water-glycerol mixtures are Newton fluids, and as such, display the behavior described in section 3.1.1. The critical frequency is determined experimentally and plotted as a function of $L^*$. The calibration results for the wires displayed in Fig. 3, consisting of 6.8 nm $\gamma$-$Fe_2O_3$ particles, are provided in Table 3, resulting in $\Delta\chi = \chi^2/(2 + \chi)$ = 0.056 ± 0.006 and $\chi$ = 0.36 ± 0.03. Recent studies have shown that by increasing the particle size of $\gamma$-$Fe_2O_3$ from 7 to 13 nm [19], we can significantly enhance the anisotropy of susceptibility $\Delta\chi$, reaching values 100 times larger than those previously obtained. In this case, the magnetic torque applied to the wire is higher and fluids with viscosities up to 100 Pa s can be studied [4,31].

| Complex fluids studied | Comparison rheometry | Quantities measured | Rheological class of materials | Reference |
|---|---|---|---|---|
| Water-Glycerol mixtures | Yes | $\eta$ | Newtonian liquid | [21,29] |
| Surfactant wormlike micelles | Yes | $\eta, G, \tau$ | Maxwell fluid | [30] |
| Polysaccharide gels | Yes | $G_{eq}, G$ | Soft solid | [23] |
| Cytoplasm of living cells | Not accessible* | $\eta, G, \tau$ | Viscoelastic liquid | [19] |
| Pulmonary surfactant w/o nanoparticles | Yes | $\eta, G_{eq}, G$ | Newtonian liquid – Soft solid | [29,31] |
| Human pulmonary mucus | Not accessible* | $\eta, G, \tau$ | Viscoelastic liquid | [32] |
| DNA and NAP proteins | Not accessible* | $\eta, G_{eq}, G$ | Newtonian liquid – Soft solid | [4] |

*Table 2*: List of synthetic and biological complex fluids studied in our group using the Magnetic Rotational Spectroscopy (MRS) method. These fluids comprise water-glycerol mixtures [21], surfactant wormlike micelles [30], polysaccharide gels [23], the cytoplasm of living mammalian cells [19], a pulmonary surfactant mimetic having been exposed to nanoparticles [31], human pulmonary mucus [32] and the system discussed in this chapter, DNA and Hfq-CTR proteins [4].

| Glycerol weight concentration (wt. %) | Static viscosity (mPa s) | $\Delta\chi$ | $\chi$ |
|---|---|---|---|
| 49.8 | 4.97 | 0.058 | 0.37 |
| 77.6 | 35.5 | 0.054 | 0.36 |
| 93.6 | 270.9 | 0.058 | 0.37 |
| Average | | 0.056 ± 0.006 | 0.36 ± 0.03 |

*Table 3*: Determination of the anisotropy of the magnetic susceptibility $\Delta\chi$ featuring in Eqs. 2, 3 and 5 for wires embedded in water-glycerol mixtures of different viscosities and studied by MRS.

## 3.3 Sample Preparation

A volume of 0.5 µL containing $10^5$ wires in phosphate buffer saline PBS (pH 7.4) is added to 100 µL of DNA, Hfq-CTR or DNA/Hfq-CTR dispersions and gently stirred (*see* **Note 9**). A volume equal to 25 µL of this dispersion is then deposited on a glass plate and sealed into a Gene Frame® (*see* **Note 6**). The glass plate is then introduced into the device generating a rotational magnetic field and investigated after the sample reaches a stationary temperature.

### 3.4 Microrheology of DNA and protein dispersions
### 3.4.1 DNA

Fig. 6a illustrates the rotation of a 33 µm wire in a 7.3 mM (dA:dT)$_{59}$ DNA dispersion at angular frequency $\omega$ = 0.31 rad s$^{-1}$, with a counter-clockwise rotation in the direction of the yellow arrows. The different images of the chronophotograph are taken at time intervals of 2 s during a 180°-rotation of the object (*see* **Note 10**). Fig. 6b displays the corresponding time dependence of the angle $\theta(t)$ at this frequency. There, $\theta(t)$ increases linearly with time according to $\theta(t) = \omega t$, indicating that the wire rotates synchronously with the field. By increasing the frequency to 3.1 rad s$^{-1}$ (Fig. 6c) and 9.4 rad s$^{-1}$ (Fig. 6d), the wire presents a transition to asynchronous rotation whereby $\theta(t)$ shows back-and-forth oscillations. The average rotational velocity $\Omega$ is indicated on the figures by straight lines in red. At the angular frequency of 9.4 rad s$^{-1}$, Fig. 6f shows that after an initial increase in the orientation angle in the counter-clockwise direction (yellow arrows), the wire undergoes a back motion in the clockwise direction (red arrow) followed again by an increase in orientation angle. Such a motion is characteristic of the asynchronous regime. For the wire in Fig. 6, $\omega_C$ is equal to 0.9 ± 0.1 rad s$^{-1}$. Repeated on wires of different lengths, the previous experiment allows us to establish a set of the critical frequencies $\omega_C$ and to adjust these data with the law in Eq. 3 leading to $\omega_C \sim 1/L^{*2}$. This approach results in a more accurate determination of the static viscosity, here found at $\eta_{DNA}$ = 1.6 ± 0.3 mPa s [4]. Such a value is slightly larger than that of the solvent, and in agreement with the Einstein law for dilute non-interacting colloids [28].

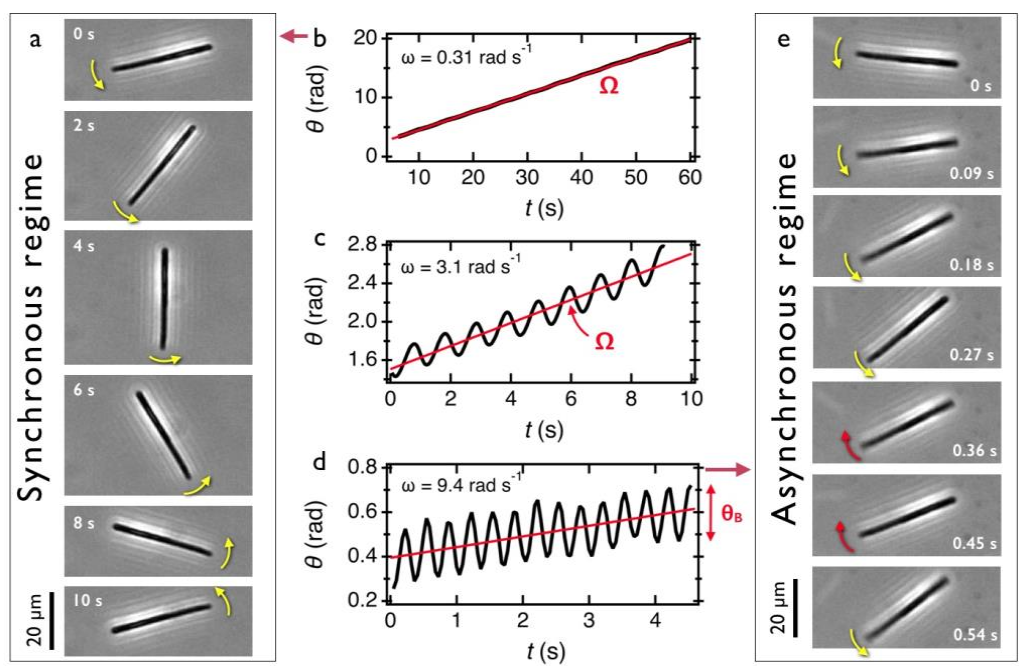

*Figure 6*: *a) Optical microscopy images of a 33 µm wire in DNA 7.3 mM subjected to a rotating field of 10.3 mT at the frequency of 0.31 rad s$^{-1}$ (T = 25 °C). The arrows indicate a steady counter-clockwise rotation. b) Time dependence of the angle $\theta(t)$ corresponding to the experiment in a). c and d) Same representation as in b) for angular frequencies 3.1 and 9.4 rad s$^{-1}$, respectively. These frequencies are in the non-synchronous regime where the wire displays back-and-forth oscillations. The critical frequency for this wire is $\omega_C$ = 0.9 rad s$^{-1}$. e) The microscopy images show that after a counter-clockwise rotation, the wire comes back rapidly in a clockwise motion, indicating that the wire rotation is hindered and in the asynchronous regime.*

### 3.4.2 Proteins

A 1.8 mM dispersion of CTR was studied using wires of different lengths by increasing the angular frequency $\omega$ from 0.03 to 30 rad s$^{-1}$ (*see* **Note 10**). For the wires followed by optical microscopy, a critical frequency between a synchronous and asynchronous regime was observed, indicating that the protein dispersion is a Newton viscous liquid. It was found in addition that $\omega_C$ varied as $1/L^{*2}$, which allowed us to determine the CTR dispersion viscosity, using Eq. 3. The CTR dispersion viscosity is found equal to $\eta_{CTR}$ = 3.3 ± 0.9 mPa s, that is about four times the viscosity of water. The slight increase of the viscosity compared to that of DNA could be due to the formation of amyloid fibers [40]. The molar concentration, the prefactor in Eq. 3 and the static shear viscosity of the single component dispersions (DNA(AT59), Hfq-CTR and the mutant Hfq-CTR$_{Mut}$) are listed in Table 4. The complete set of data for DNA and Hfq-CTR dispersions, including the demonstration of the law described by Eq. 3 as a function of the reduced length $L^*$ can be found in reference [4]

| DNA-protein sample | Concentration (mM) | Prefactor in Eq. 3 (rad s$^{-1}$) | Viscosity $\eta$ mPa s | $\Delta\eta$ mPa s |
|---|---|---|---|---|
| DNA(AT59) | 7.3 | 1061 ± 90 | 1.6 | 0.3 |
| Hfq-CTR | 1.82 | 519 ± 74 | 3.3 | 0.9 |
| Hfq-CTR$_{Mut}$ | 1.82 | 723 ± 13 | 2.4 | 0.3 |

**Table 4**: Determination of the static viscosity for DNA, Hfq-CTR and DNA/Hfq-CTR complexes showing viscous behavior.

### 3.5 Microrheology of DNA-Amyloid complexes

DNA/Hfq-CTR complexes at molar concentrations 1.8/1.8, 7.8/3.9 and 7.3/1.8 were investigated with the MRS technique. Magnetic wires of length between 13 and 80 µm were activated by a rotating field operating at angular frequency of 0.03 to 30 rad s$^{-1}$. For DNA/Hfq-CTR complexes, measurements displayed a noticeable change in the wire oscillatory behavior as compared to that of the individual constituents, either DNA or Hfq-CTR. Here the wires oscillate between two orientations, which are in addition independent of the angular frequency. Fig. 7a display chronophotographs of a 19.9 µm wire (diameter 2.0 µm) incorporated in a 4:1 DNA/Hfq-CTR complex and actuated at $\omega$ = 0.031 rad s$^{-1}$. The background image under phase contrast microscopy appears heterogeneous, which could come from Hfq-CTR assembled structures on DNA with sizes in the visible light range. For this sample, wire rotation is found to be asynchronous regardless of the frequency. Translated in terms of orientation angle $\theta(t)$, the time dependence displays regular oscillations characterized by a zero average rotational velocity $\Omega(\omega)$. Similar results are obtained at higher frequencies, $\omega$ = 0.31, 3.1 and 31 rad s$^{-1}$, as illustrated in Fig. 7c, 7d and 7e, respectively. At the angular frequency of 31 rad s$^{-1}$, Fig. 7f shows that after an initial increase in the orientation angle in the counter-clockwise direction (yellow arrows), the wire undergoes a back motion in the clockwise direction (red arrow) followed again by an increase in orientation angle. Except for the amplitude of the oscillations which decreases with frequency, this movement is identical to that observed three decades in frequency below (Fig. 7a). The observation on DNA/Hfq-CTR is an indication that the material is a soft solid and characterized by a yield stress behavior (Fig. 5 and Table 1). For this particular experiment, the low and high frequency oscillation angles can be deduced, leading to an estimate of the elastic moduli $G_{eq}$ and $G$. Repeated on wires of different lengths, the previous experiment allows us to establish a statistic set of angles $\theta_{eq}$ and $\theta_0$ and to adjust these data with the law in Eq. 5b [4]. This approach results in a more accurate determination of the moduli, $G_{eq}$ = 1.7 ± 0.4 Pa and $G$ = 3.6 ± 0.9 Pa. The complete set of data for DNA/Hfq-CTR complexes, including the demonstration of the law described by Eq. 5 as a function of the reduced length $L^*$ can be found in reference [4]. In conclusion, the 4:1 DNA/Hfq-CTR complex shown here behaves as a soft gel with a clear viscoelastic signature. The gel formation is here attributed to the crosslinking of Hfq-CTR amyloid fibers with DNA molecules. This outcome might have implications for gene expression regulation and other cellular processes involving genetic transport in bacteria.

In the previous section, we have shown that by simple measurements of the behavior of magnetic wires subjected to a rotating field, the rheological nature of the fluids or materials studied could be identified. The observation of a soft solid type behavior for DNA and Hfq-CTR protein dispersions is a remarkable result as it suggests the formation of a network, probably of amyloid fibers cross-linked with the DNA strands. Moreover, with the quantitative method described above, we can determine the viscosity and elasticity values of the studied systems, which represents an important step in the comparison with theoretical models of complex fluid dynamics [28,40].

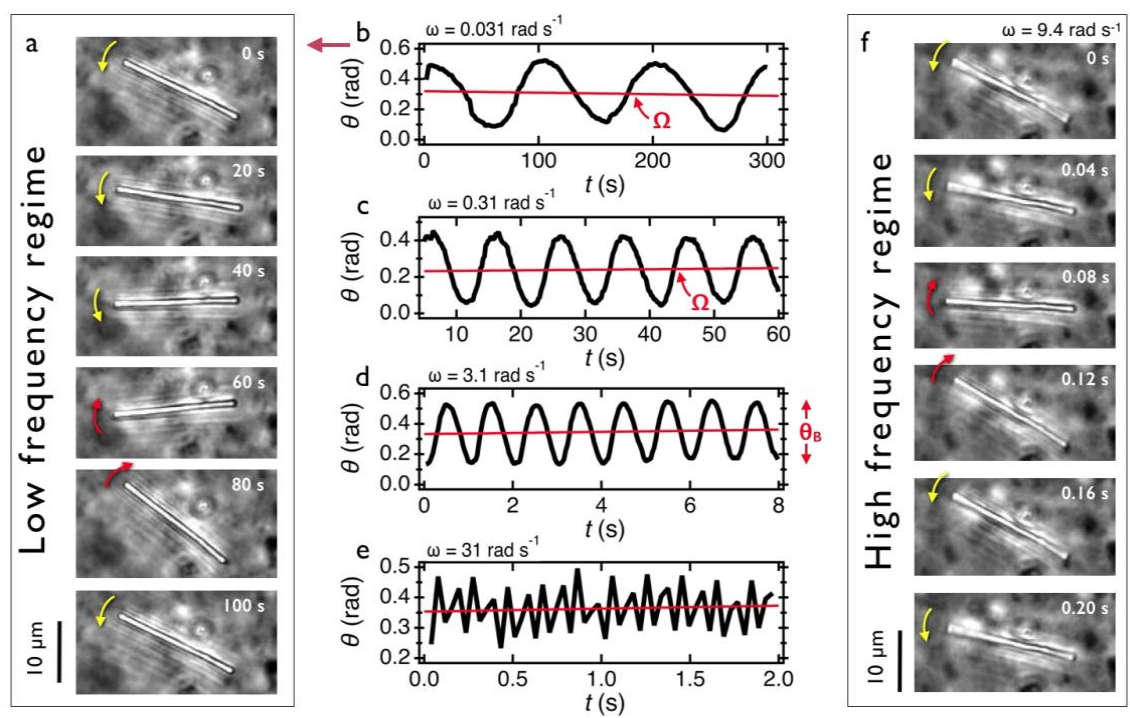

*Figure 7*: a) Optical microscopy images of a 19.9 µm wire in 4:1 DNA /Hfq-CTR dispersion subjected to a rotating field of 10.3 mT at the frequency of 0.031 rad s$^{-1}$ (T = 25 °C). After a steady counter-clockwise rotation (yellow arrows), the wire comes back rapidly in a clockwise motion (red arrows), indicating that the wire rotation is hindered and in the asynchronous regime. b, c, d and e) Time dependence of the angle $\theta(t)$ corresponding to experiments carried out at angular frequencies 0.031, 0.31, 3.1 and 31 rad s$^{-1}$, respectively. f) As in Fig. 7a for 9.4 rad s$^{-1}$.

**4 Notes**

1. Under the preparation conditions of the 20 g L$^{-1}$ Hfq-CTR peptide stock solution, Hfq-CTR spontaneously self-assembles into large amyloid fibers on time scales of several days [41]. For the preparation of the DNA/Hfq-CTR mixed solution, it is verified that the proteins are not already assembled into amyloid fibers.

2. The NH$_4$Cl salt is used for screening the electrostatic interaction between the PDADMAC and γ-Fe$_2$O$_3$@PAA$_{2K}$ during the synthesis of the magnetic wires. Dialysis is performed using a 1M NH$_4$Cl dispersion of PDADMAC and γ-Fe$_2$O$_3$@PAA$_{2K}$ at concentration 1 g L$^{-1}$ against ultrapure MilliQ water. During dialysis, the ionic strength is decreased from 1 M to 10$^{-3}$ M, leading to the formation of micrometer-sized iron oxide wires (see Section 2.2).

3. Concerning the rotating field device shown in Fig. 1, although the same voltage is applied to the two pairs of coils, differences between the X and Y components of the magnetic field may be present. In such a case, the field is elliptical, leading to inaccuracies in the determination of $\omega_C$, and of the viscosity. The voltages in each pair of coils must then be adjusted to make the fields in both directions equal.

4. Electromagnets, as the ones used in MRS experiments, have the advantage that the magnetic field is easily adjusted through the voltage applied to the coils. The disadvantage of this configuration is however that the currents in the coils can be large, typically above 1 A, leading to an overheating of the coils, and of the sample environment. For such experiments, it is necessary to control the thermal environment of the device, which can be ensured with a temperature-controlled stream of air directed toward the measuring cell (Fig. 1).

5. Relevant frequency ranges in identifying the rheological nature of a fluid are generally in the low and high frequency ranges. Low frequencies refer here to $\omega$ of the order of 10$^{-3}$ rad s$^{-1}$ and below. At this frequency, the time for a wire to make a full rotation is of the order of two hours, which is only possible if the experiment being conducted is stable over time. The stability here concerns the position of the wire in the sample, the temperature, or the tightness of the cell containing the sample. In the high frequency regime for the MRS experiment, limitations are the exposure time and time data transfer between the QImaging EXi Blue camera and the computer. Taking images at a rate higher than 100 frames per second requires the use of an ultrafast camera.

6. The samples studied in microrheology are placed between a glass slide and a cover slide, with the Gene Frame serving as a spacer between the two. The Gene frame interior dimensions are 10×10×0.25 mm$^3$, corresponding to a volume of 25 μL of fluid.

7. The ImageJ plugin used for tracking the 2D-motion of wires computes for each image their center of mass, length and orientation angle with an arbitrary reference. We use the orientation angle $\theta(t)$ and its time derivative $\Omega = \langle d\theta(t)/dt \rangle_t$ as relevant quantities in the interpretation of the DNA and Hfq-CTR data. The angular resolution in $\theta(t)$ decreases with the length as $2 \times 10^{-12} L^{-1.76}$ [42]. This expression gives an angular resolution of 1.3° for a 2 μm wire and 0.02° for a 20 μm wire.

8. When possible, it is also imperative to confirm the data obtained by microrheology by the standard rheometry measurements. Given that the types of flows in rheometry and microrheology are not the same, this comparison is all the more necessary and allows to realize the precision of the technique implemented.

9. During sample preparation, it is important to add a minimal amount of the wire dispersion (< 0.5 µL) to keep the rheological properties of the sample unchanged. This is a recurring issue in microrheology using tracking particles. Note also that with complex fluids of high viscosity (> 100 Pa s), mixing the wires with the sample becomes difficult, and requires adapted protocols.

10. It is also crucial with MRS to control the density of the wires within the sample. In particular, wires should not be too close to each other to minimize their mutual interactions. Estimates of velocity fields around a rotating wire in water show that the distance between two wires should be greater than 5 times the length of the longer wire to reduce hydrodynamic interaction. Such a distance is in general sufficient to minimize magnetic dipolar interactions, which could also hinder their rotation.